\documentclass[runningheads]{llncs}
\usepackage[T1]{fontenc}
\usepackage{graphicx}
\usepackage{soul}
\usepackage{url}
\usepackage[hidelinks]{hyperref}
\usepackage[utf8]{inputenc}
\usepackage[small]{caption}
\usepackage{amsmath}
\usepackage{booktabs}
\usepackage{algorithm}
\usepackage{algorithmic}
\usepackage{stfloats}
\usepackage{tikz}
\usepackage{multirow}
\usepackage{subfigure}
\usepackage{enumitem}

\newtheorem{dfn}{Definition}
\newcommand\paratitle[1]{\smallskip\noindent\textbf{#1}}
\usepackage{tikz}
\usetikzlibrary{automata, positioning, arrows}
\tikzset{
->,
>=stealth', 
node distance=1.6cm, 
every state/.style={thick, fill=gray!10, minimum size=0pt}, 
initial text=$ $ 
}

\begin{document}

\title{The Burden of Being a Bridge:
Analysing Subjective Well-Being of Twitter Users during the COVID-19 Pandemic}

\author{Anonymous author(s)}
\author{
    Ninghan Chen\textsuperscript{\rm 1},
    Xihui Chen\textsuperscript{\rm 2},
    Zhiqiang Zhong\textsuperscript{\rm 1}
    Jun Pang\textsuperscript{\rm 1,2}\\
}
\institute{
\textsuperscript{\rm 1} Faculty of Sciences, Technology and Medicine, \\ University of Luxembourg, L-4364 Esch-sur-Alzette, Luxembourg\\
\textsuperscript{\rm 2} 
Interdisciplinary Centre for Security, Reliability and Trust, \\ University of Luxembourg, L-4364 Esch-sur-Alzette, Luxembourg
}

\titlerunning{The burden of being a bridge}
%
%
%
\maketitle              
\begin{abstract}
The outbreak of the COVID-19 pandemic triggers \emph{infodemic} over 
online social media, which  significantly impacts public health around the world, 
both physically and psychologically. 
In this paper, we study the impact of the pandemic on the mental health of 
influential social media users, whose sharing behaviours significantly promote 
the diffusion of COVID-19 related information. 
Specifically, we focus on subjective well-being (SWB), 
and analyse whether SWB changes have a relationship with their 
\emph{bridging performance} in information diffusion, 
which measures the speed and wideness gain of information transmission due to their 
sharing. We accurately capture users' bridging performance by proposing a new measurement. 
Benefiting from deep-learning natural language processing models, we quantify 
social media users' SWB from their textual posts. 
With the data collected from Twitter for almost two years,
we reveal the greater mental suffering of influential users during 
the COVID-19 pandemic. 
Through comprehensive hierarchical multiple regression analysis,
we are the first to discover the strong {relationship} between social users' SWB and 
their bridging performance.


\end{abstract}

\section{Introduction}
\label{sec:introdution}
Since its outbreak, COVID-19 has become an unprecedented global health 
crisis and incited a worldwide \emph{infodemic}. 
The term ``infodemic'' outlines the perils of misinformation 
during disease outbreaks mainly on social media~\cite{Cielli2020Infodemic,Guarino2021infodemic}.
Apart from accelerating virus transmission by distracting 
social reactions, 
the infodemic increases cases of psychological diseases such as anxiety, phobia 
and depression during the pandemic~\cite{dubey2020psychosocial}. 
As a result, the infodemic impairs the UN's sustainable development goals (SDGs), 
especially SDG3 which aims to promote mental health and 
well-being.

To combat infodemic, both governments and healthcare bodies  
have launched a series of social media campaigns to diffuse  
trustworthy information. 
To amplify the speed and wideness of information 
spread, users with large number of followers are invited to  
help transmit messages~\cite{zarocostas2020fight,BM21,yang2021covid}. 
Healthcare professionals and social activists
also voluntarily and actively take part in relaying information they 
deem as useful with their social media accounts. 
All these people actually play a bridging role on social media 
delivering information to the public, although their \emph{bridging performance} differs.
We use bridging performance as an analogy to estimate 
how efficient and wide information can spread across social 
media due to the sharing of a user. 

Subjective well-being (SWB), one important indicator of SDG3, 
evaluates individuals’ cognitive (e.g., life satisfaction)
and affective (i.e., positive and negative) perceptions of 
their lives~\cite{JaidkaPNAS20}.
Since the onset of the COVID-19 pandemic,
the decrease of SWB has been unanimously recognised across the world.
With studies for various sub-populations~\cite{HLCX20,ENWswb21}, 
many factors have been discovered correlating to SWB changes such as professions,
immigration status and gender. 
In this paper, we concentrate on influential social media users who play the bridging
role in diffusing COVID-19 information, and 
study the impact of the pandemic on their SWB. 
We further examine whether their active participation in diffusing 
COVID-19 information is a predictor of the SWB changes.  
To the best of our knowledge, we are the first to study the mental 
health of this specific group of people during the pandemic. 

We identify two main challenges to overcome before conducting our analysis. 
First, there are no measurements available that can accurately quantify
users' real bridging performance in diffusing COVID-19 related information.
The measurements, widely used in crisis communications and online marketing, 
rely on social connections, and have been found insufficient in capturing 
users' actual bridging performance, especially 
in such a global health crisis~\cite{struweg2020twitter}.
For instance, although some healthcare professionals are not super tweeters with thousands 
of followers, their professional endorsement significantly promotes the 
popularity of the posts they retweeted~\cite{struweg2020twitter}.
The second challenge is the access to the SWB levels of a large 
number of social media users whose bridging performance is simultaneously available.  

In this paper, we take advantage of the information outbreak on social media incurred by 
the COVID-19 pandemic and the advances of artificial intelligence
to address the two challenges.
For the first challenge, we propose a new bridging performance measurement 
based on \emph{information cascades}~\cite{CZTZZ19,WSLGC17}
which abstract both information spread processes and social connections.
To address the second challenge, we leverage the success of 
deep learning in Natural Language Processing (NLP) and estimate users' 
SWB by referring to the sentiments expressed in their textual posts.
In spite of the inherent biases, the power of social media posts 
has been shown in recent studies~\cite{jaidka2020estimating} for 
robust extraction of well-being with supervised data-driven methods. 
In this paper, instead of manually constructed features, 
we use the state-of-the-art transformer-based text embedding to 
automatically learn the representative features of textual posts.

\paratitle{Our contributions.} 
We collect data from Twitter generated from  
\emph{the Greater Region of Luxembourg} (GR).  %
GR is a cross-border region centred around Luxembourg and composed of 
adjacent regions of Belgium, Germany and France. 
One important reason to select this region is its intense inter-connections of 
international residents from various cultures, which is unique 
as a global financial centre.  
Moreover, they well represent the first batch of countries administering COVID vaccines. 
Our collection spans from October 2019 to the end of 2021 for over 2 years,
including 3 months before the outbreak of the COVID-19 pandemic.
Our contributions are summarised as follows:
\begin{itemize}[leftmargin=*]
\item We propose a new measurement to capture the actual bridging performance 
of individual users in diffusing COVID-19 related information.
Compared to existing social connection-based measurements, it is directly derived from 
information diffusion history. 
Through manual analysis of the collected dataset, our measurement allows for 
identifying the accounts of influential health professionals and volunteers 
that are missed before in addition to super tweeters. 
\item Through deep learning-based text embedding methods, we implement 
a classification model which can accurately extract 
the sentiments expressed in social media messages.
With the sentiments of posts, we quantitatively estimate individual users' SWB, 
and confirm the greater suffering of influential users in their SWB
during the pandemic. 
\item Through the hierarchical multiple regression model, we reveal that  
users' SWB has a strong negative {relationship} with their bridging performance in COVID-19
information diffusion, but weak {relationship} with their social connections.
\end{itemize}
%

Our research provides policy makers with an effective method to 
identify influential users in the fight against infodemic. 
Moreover, we contribute to the realisation of SDG3 by highlighting 
the necessity to pay special attention to the mental well-being 
of people who actively participate in transmitting information in health crises 
like COVID-19.
\section{Related Work} 
\label{sec:Related work}
\subsection{Measuring bridging performance}
A considerable amount of literature has been published quantifying users' bridging performance 
based on social connections to identify amplifiers in social media.
We can divide the measurements into two types.
The first type of measurements implicitly assume that influential users are likely to hold certain 
topology properties on social networks such as large degrees, strong betweenness centrality or 
community centrality~\cite{freeman1978centrality}.
The second type of measurements assume that influential users tend to be more likely reachable 
from other users through random walks.
PageRank~\cite{page1999pagerank} and its variant TwitterRank~\cite{weng2010twitterrank}
among the representative benchmarks of this type of measurements. 
PageRank is calculated only with network structures while 
TwitterRank additionally takes into account topic similarities between users.
All the two categories of measurements have been widely applied in practice, 
from public health crisis communication~\cite{mirbabaie2020social} to online 
marketing~\cite{li2011discovering}. 
However, recent studies pointed out that they may not truly capture users' actual bridging performance 
in information diffusion during a specific public healthy crisis~\cite{huang2014identifying,mirbabaie2020social}. 
Although new measurements are proposed by extending existing ones with fusion indicators, 
their poor efficiency prevents them from being applied to real-world large-scale networks 
like Twitter and Facebook~\cite{huang2014identifying}.



\subsection{Subjective well-being extraction}
Subjective well-being is used to measure how people subjectively rate their lives 
both in the present and in the near future~\cite{diener2003personality}. 
Many methods have been used to assess subjective well-being, from 
traditional self-reporting methods~\cite{diener1985satisfaction} to
the recent ones exploiting social media~\cite{yang2016life,liu2015facebook}.
Studies have cross-validated SWB extracted from social media data with the Gallup-Sharecare 
Well-Being Index survey,\footnote{https://www.gallup.com/175196/gallup-healthways-index-methodology.aspx} 
a classic reference used to investigate public SWB, and found that SWB extracted from 
social media is a reliable indicator of SWB~\cite{jaidka2020estimating}. 
Twitter-based studies usually calculate SWB as  
the overall scores of positive or negative emotions (i.e., sentiment 
or valence)~\cite{JaidkaPNAS20}.  
Sentiment analysis has developed from the original lexicon-based 
approaches~\cite{DHBPlos11,bradley1999affective} to the data-driven ones 
which ensure better performance~\cite{JaidkaPNAS20}.
We adopt the recent advances of the latter approaches, and make use of 
the pre-trained XLM-RoBERTa~\cite{OL20}, a variant of RoBERTa~\cite{liu2019roberta}, to 
automatically learn the linguistic representation of textual posts. 
As a deep learning model, RoBERTa and its variants have been shown to 
overwhelm traditional machine learning models in capturing the linguistic patterns of 
multilingual texts~\cite{BCEN20}.

\section{The GR-ego Twitter Dataset}\label{sec:data}
In this section, we describe how we build our Twitter dataset, 
referred to as \emph{GR-ego}. In addition to its large number of active users, 
we have another two considerations to select Twitter as our data source. 
%
First, the geographical addresses of posters are attached with tweets and 
thus can be used to locate users.
Second, tweet status indicates whether a tweet is retweeted. 
If a tweet is retweeted,  the corresponding original tweet ID is provided. 
Together with the time stamps, we can track the diffusion process of an original tweet.   
Our GR-ego dataset consists of two components: 
(i) the social network of GR users recording 
their following relations;
(ii) the tweets posted or retweeted by GR users during the pandemic. 
We follow three sequential steps to collect our GR-ego dataset. 
Table~\ref{table:summary_GR} summarises its main statistics. 

\paratitle{Step 1. Meta data collection.} 
Our purpose of this step is to collect seed users in GR who actively
participate in COVID-19 discussions. 
Instead of directly searching by COVID-19 related keywords, 
we make use of a publicly available dataset 
of COVID-19 related tweets for the purpose of efficiency~\cite{COVID-19Dataset}. 
Restricted by Twitter's privacy policies, this dataset only consists of 
tweet IDs. We extract the tweet IDs posted between October 22nd, 2019, which
is about three months before the claim of the COVID-19 pandemic, and 
December 31st, 2021. 
Then with these IDS, we download the corresponding tweet content. 
On Twitter, geographical information, i.e., the
locations of tweet posters and original users if tweets are re-tweeted, 
is either maintained by Twitter users, or provided directly by their 
positioning devices.
We stick to the device-input positions, and 
only use user-maintained ones when such positions are unavailable. 
Due to the ambiguity of user-maintained positions,
we leverage the geocoding APIs, Geopy 
and  ArcGis Geocoding 
to regularise them into machine-parsable locations. 
%
With regularised locations, we filter the crawled tweets and only retain 
those from GR.  In total, we obtain 128,310 tweets from 8,872 GR users. 

\paratitle{Step 2. Social network construction.}
In this step, we search GR users from the seed users and 
construct the GR-ego social network. 
We adopt an iterative approach to gradually enrich the social network.
For each seed user, we
obtain his/her followers and only retain those who have a mutual following
relation with the seed user, 
because such users are more likely to reside in
GR. We then extract new users' locations from their profile data 
and regularise them. 
Only users from GR are added to the social network as new nodes.
New edges are added if there exist users in the network with following {relationships}
with the newly added users. 
%
After the first round, we continue going through the newly added users 
by adding their mutually followed friends that do not exist in the 
current social network. This process continues until no new users can be 
added. Our collection takes $5$ iterations before termination.
In the end, we take the largest weakly connected component 
as the \emph{GR-ego social network}.

\begin{table}[!t]
\caption{ Statistics of the GR-ego dataset.}
\label{table:summary_GR}
\centering
{
\begin{tabular}{|p{15mm}|p{48.5mm}|r|}
\hline
\multirow{3}{15mm}{Social network} & \#node                                   & 5,808,938  \\ \cline{2-3} 
                                & \#edge                                   & 12,511,698 \\ \cline{2-3} 
                                & Average degree                          & 2.15       \\ 
                                \hline
\multirow{5}{15mm}{Timeline tweets} & \#user                                    & 14,756     \\ \cline{2-3} 
                                & \#tweet before COVID                 & 5,661,949  \\ \cline{2-3} 
                                & \#tweet during COVID                 & 18,523,099 \\ \cline{2-3} 
                                & \#tweet per user before COVID & 388.44     \\ \cline{2-3} 
                                & \#tweet per user during COVID & 1255.29    \\ \hline
\end{tabular}
}
\vspace{-5mm}
\end{table}

\paratitle{Step 3. COVID-19 related timeline tweets crawling.} In this step, we collect tweets
originally posted or re-tweeted by the users in our dataset. 
These tweets will be used to extract users' SWB.  
Thus, the collected tweets are \emph{not limited} to those relevant to the COVID-19 pandemic. 
Due to the constraints of Twitter, it is not tractable to download all the users' past tweets.  
We select a sufficiently large number of representative users 
who actively participated in retweeting COVID-19 related messages,
and then crawl their history tweets. 
In detail, 
we choose 14,756 users who (re)tweeted at least three COVID-19 related messages.
With the newly released Twitter API which allows for downloading 
$500$ tweets of any given month for each user, we collect 
$24,185,048$ tweets spanning between October 22nd, 2020 and December 31st, 2021. 
This period also contains the last three months before the pandemic is officially claimed.

\section{Data Processing}
\label{sec:data processing}

\subsection{Cascade computation}
\label{ssec:cascade computation}
A cascade records the process of 
the diffusion of a message. It stores all activated users and 
the time when they are activated. 
In our dataset, a user is activated in diffusing a message when he/she 
retweets the message. 
In this paper, we adopt the widely accepted \emph{cascade tree} to represent 
the cascade of a message~\cite{WSLGC17}. 

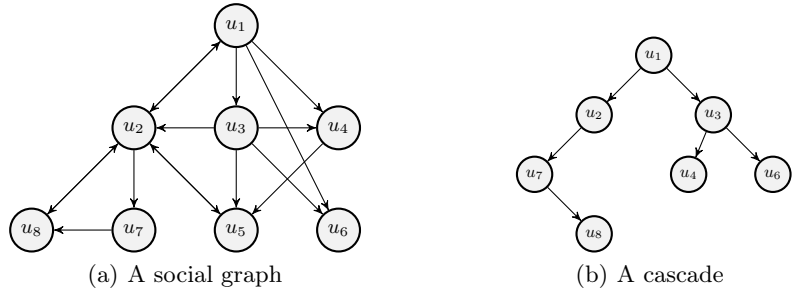
\begin{figure}[!t]
    \centering
    \subfigure[A social graph\label{fig:social graph}]{
    \begin{tikzpicture}[scale = 0.85, transform shape]
        \node[state](2) {$u_2$};
        \node[state, right of = 2](3) {$u_3$};
        \node[state, below of = 2](7) {$u_7$};
        \node[state, right of = 3](4) {$u_4$};
        \node[state, above of = 3](1){$u_1$};
        \node[state, below of = 3](5) {$u_5$};
        \node[state, right of = 5](6) {$u_6$};
        \node[state, left of = 7](8) {$u_8$};
        \draw (1) edge (2)
              (2) edge (1)
              (1) edge (3)
              (1) edge (4)
              (3) edge (5)
              (2) edge(5)
              (5) edge (2)
              (3) edge (4)
              (4) edge (5)
              (3) edge (6)
              (1) edge (6)
              (2) edge (7)
              (2) edge (8)
              (8) edge (2)
              (7) edge (8)
              (3) edge (2) ;
    \end{tikzpicture}
    }
    \hspace{5em}
   \subfigure[A cascade \label{fig:cascase example}]{
     \begin{tikzpicture}[scale = 0.7, transform shape]
        \node[state](2) {$u_2$};
        \node[state, above right of = 2](1){$u_1$};
        \node[state, below right of = 1](3) {$u_3$};
        \node[state, below right of = 3](6) {$u_6$};
        \node[state, left of = 6](4) {$u_4$};
        \node[state, below left of = 2](7) {$u_7$};
        \node[state, below right of = 7](8) {$u_8$};
        \draw (1) edge (2)
              (1) edge (3)
              (2) edge (7)
              (7) edge (8)
              (3) edge (4)
              (3) edge (6);
    \end{tikzpicture}
    }
    \vspace{-3mm}
    \caption{Example of a cascade. \label{fig:cascase} }
    \vspace{-5mm}
\end{figure}

The first user who posted the message is regarded as the root of the cascade tree. 
Users who retweeted the message, but received no further retweeting 
comprise the leaf nodes. Note that a tweet with the quotation to another 
tweet is also considered as a retweet of the quoted message. 
An edge from $u$ to $u'$ is added to the cascade if $u'$ follows $u$ and 
$u'$ re-tweeted the message after $u$, indicating $u$ activated 
$u'$. If many of the users who $u'$ follows ever retweeted 
the message, meaning $u'$ may be activated by any of them, 
we select the one who lastly retweeted as the parent node of $u'$.
Figure~\ref{fig:cascase example} shows a cascade of the social network 
in Figure~\ref{fig:social graph}. 
In this example, user $u_4$ can be activated by the messages
retweeted by either $u_1$ or $u_3$. Since $u_3$ retweeted after $u_1$, we add the 
edge from $u_3$ to $u_4$ indicating that the retweeting of $u_3$ activated $u_4$.

We denote the root node of a cascade $C$  by $r(C)$.
We call a path that connects the root and a leaf node a \emph{cascade path}, 
which is actually a sequence of nodes ordered by their activation time. 
For instance, $(u_1, u_3, u_4)$ is a cascade path in our example indicating that
the diffusion of a message started from $u_1$ and reached $u_4$ in the end through
$u_3$. In this paper, we represent a cascade tree as a set of cascade paths. 
For instance, the cascade in Figure~\ref{fig:cascase example} is represented 
by the following set $\{ (u_1, u_2, u_7, u_8), (u_1, u_3, u_4), (u_1, u_3, u_6)\}$.

For our study, we follow the method in~\cite{kupavskii2012prediction} 
to construct tweet cascades. Recall that when a tweet's status is `{\it Retweeted}', 
the ID number of the original tweet is also recorded. 
We first create a set of original tweets with all the ones labelled in our meta data 
as `{\it Original}'. 
Second, for each original tweet, we collect the IDs of users that have
retweeted the message. At last, we generate the cascade for 
every original tweet based on the following relations in our 
GR-ego social network and their retweeting time stamps. 
We eliminate cascades with only two users where messages are just 
retweeted once. 
In total, 614,926 cascades are built and the average size of these cascades is 7.13. 

%

\subsection{Sentiment analysis}
\label{ssec:sentiment analysis}
Previous works~\cite{ZJZ20} leverage user-provided mood (e.g., angry, excited) 
or status to extract users' sentiment (i.e., positive or negative) and use them 
to approximately estimate affective subjective well-being.
However, such information is not available on Twitter. 
We refer to the sentiments expressed in textual posts to extract users' SWB.
In this paper, we treat sentiment extraction as a 
tri-polarity sentiment analysis for short texts,  
and classify a tweet as \emph{negative}, \emph{neutral} or \emph{positive}. 
In order to deal with the multilingualism of our dataset, 
we benefit from the advantages of deep learning in sentiment analysis~\cite{BCEN20},
and build an end-to-end deep learning model to conduct the classification.
Our model is composed of three components.
The first component uses a pre-trained multilingual language model, i.e., XLM-RoBERTa~\cite{OL20}, 
to calculate the representation of tweets. 
The representations are then sent to the second component, a fully-connected ReLU layer 
with dropout.
The last component is a linear layer added on the top of the second component’s 
outputs with sigmoid as the activation function.
We use cross-entropy as the loss function and optimise it with the Adam optimiser.

\smallskip\noindent
\textbf{Model training and testing.}
We train our model on the \emph{SemEval-2017 Task 4A} 
dataset~\cite{rosenthal2017semeval}, which has been used for sentiment analysis 
on COVID-19 related messages~\cite{duong2020ivory,nguyen2020bertweet}.
The dataset contains 49,686 messages which are annotated with one of the three
labels, i.e., positive, negative and neutral.  
We shuffle the dataset and take the first 80\% for training and 
the rest 20\% for testing. 
We assign other training parameters following the common principles in existing works. 
We run 10 epochs with the maximum string length as 128 and dropout ratio as 0.5. 
When tested with macro-average F1 score and accuracy metrics, 
we achieve an accuracy of 70.09\% and macro-average F1 score of 71.31\%.
\begin{figure}[t]
    \centering
     \includegraphics[scale = 0.6]{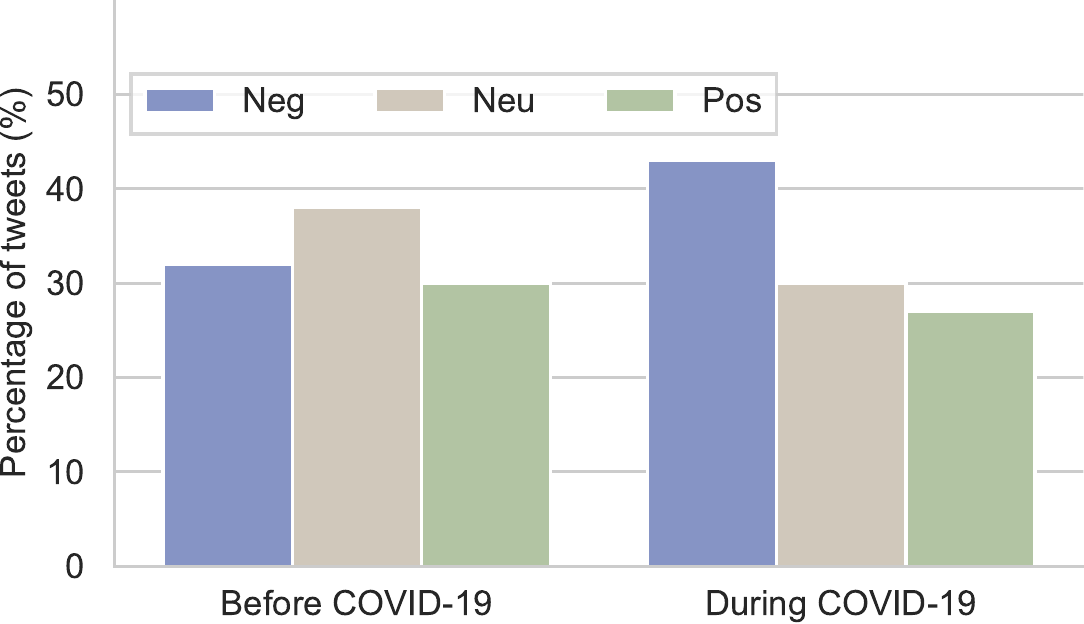}
     \caption{Sentiment distribution of users' timeline tweets.}
     \label{fig:sentiment dist}
     \vspace{-5mm}
 \end{figure}

In spite of its effectiveness on classifying \emph{SemEval-2017 Task 4A} data, 
in order to check whether such performance will persist on our GR-ego dataset, we construct 
a new testing dataset. This dataset consists of $500$ messages, $100$ for 
each of the top 5 most popular languages. We hire two annotators to manually
label the selected tweets and the annotated labels are consistent between them with
Cohen's Kappa coefficient $k = 0.93$. When applied on this new manually annotated dataset, 
our trained model achieves a similar accuracy of about 87\%. 

\paratitle{Analysing our GR-ego dataset.}
Before applying our sentiment classification model on our GR-ego dataset, 
we clean tweet contents by removing all URLs, and mentioned usernames. 
Figure~\ref{fig:sentiment dist} summarises the statistics obtained  
from user timeline tweets before and during the pandemic.
The numbers of users' timeline tweets are consistent with previous studies. 
For instance, users tend to become more negative after the outbreak of the COVID-19 
pandemic~\cite{ENWswb21,HLCX20}.

\section{Bridging Performance of Users in Information Diffusion}
\label{sec:role}
We devote this section to addressing the first challenge regarding identifying users 
that play the bridging role in  transmitting COVID-19 related information.
Specifically, we propose a measurement for 
the bridging performance of social media users in the diffusion of COVID-19 related information.

\subsection{Measuring user bridging performance}
\label{ssec:ubm}
We evaluate users' overall performance in the
diffusion of observed COVID-19 related tweets.
As a user can participate in diffusing a number of tweets, we first focus on 
her/his importance in the diffusion of one single tweet and then combine her/his 
importance in all tweets into one single measurement. 
We consider a user \emph{more important} in diffusing a tweet when his/her 
retweeting behaviour activates more users, or leads to a 
given number of activated users with fewer subsequent retweets.
In other words, a more important user promotes wider acceptance of 
the information or accelerates its propagation. 


Given a cascade path $S\!=\!(u_1, u_2, \ldots, u_n)$, we use $S^*(u_i)$ ($1\le i< n$)
to denote the subsequence composed of the nodes after $u_i$ (including $u_i$),  
i.e., $(u_i, u_{i+1}, \ldots, u_n)$.  
For any $u$ that does not exist in $S$, we have 
$S^*(u) = \varepsilon$ where $\varepsilon$ represents an empty sequence and 
its length~$|\varepsilon|=0$. 
\begin{dfn}[Cascade bridging value] \label{def:bridgingvalue}
    Given a cascade tree $C$ and a user $u$ ($u\neq r(C)$), the cascade bridging value of 
    $u$ in $C$  is calculated as:
\begin{displaymath}
\alpha_C(u) = \left(\sum_{S\in C} \frac{\mid S^*(u)\mid}{\mid S\mid}\right)/{|C|}.
\end{displaymath}
\end{dfn}
Note that our purpose is to evaluate the importance of users as transmitters 
of messages. Therefore, the concept of cascade bridging value is not 
applicable to the root user, i.e., the message originator.
\begin{example}
In Figure~\ref{fig:cascase example},
$u_3$ participated in two cascade paths, i.e.,  
$S_1=(u_1, u_3, u_4)$ and $S_2=(u_1, u_3, u_6)$.
Thus, $S_1^*=(u_3, u_4)$ and $S_2^*=(u_3, u_6)$. 
We then have $\alpha_C(u_3) = \frac{2/3 + 2/3}{3} \approx 0.44$.
\end{example}
In Definition~\ref{def:bridgingvalue}, we do not simply use the proportion of users activated by a user
in a cascade to evaluate her/his bridging performance. 
This is because it only captures the number of activated users and ignores the
speed of the diffusion. 
Taking $u_2$ in Figure~\ref{fig:cascase example} as an example, 
according to our definition, 
$\alpha_C(u_2) = 0.25$ which is smaller than $\alpha_C(u_3)$. 
This is due to the fact that $u_2$ activated two users through two retweets
while $u_3$ only used one. 
However, if we only consider the proportion of activated users, the values of 
these two users will be the same. 

With a user's bridging value calculated in each cascade, we define   
\emph{user bridging magnitude} to evaluate her/his overall 
importance in the diffusion of a given set of observed messages.
Intuitively, we first add up the bridging values of a user in all his/her participated
cascades and then normalise the sum by the maximum number of cascades 
participated by a user.
This method captures not only the bridging value of a user in each participated cascade, 
but also the number of cascades she/he participated in. 
This indicates that, a user who is more 
active in sharing COVID-19 related information is considered more 
important in information diffusion.
\begin{dfn}[User bridging magnitude (UBM)] \label{def:IDimportance}
    Let $\mathcal{C}$ be a set of cascades on a social network and $\mathcal{U}$
    be the set of users that participate in at least one cascade in $\mathcal{C}$. 
    A user $u$'s user bridging magnitude (UBM) is calculated as:  
\begin{displaymath}    
    \omega_\mathcal{C}(u) = 
\frac{\sum_{C\in\mathcal{C}}\alpha_{C}(u)}{
\max_{u'\in \mathcal{U}}|\{C\in\mathcal{C}|\alpha_{C}(u')>0\}|}.  
\end{displaymath}
\end{dfn}
 With this measurement, we can compare the UBM values of any two given users, and 
learn which one plays a more important role in information diffusion. 




\begin{table}[!t]
\caption{Comparison of bridging performance with benchmarks.}
\label{tab:comparison}
\centering
\resizebox{1.\linewidth}{!} 
{
\begin{tabular}{|l|r|r|r|r|r|r|}
\hline
& {in-degree}
& {PageRank} & {TwitterRank} & \begin{tabular}[c]{@{}l@{}}{Betweenness}\\ 
{centrality}\end{tabular} & \begin{tabular}[c]{@{}l@{}}{Community}\\ {centrality}\end{tabular} 
& ~{UBM}~                        \\ \hline\hline
Avg. \#activated user/minute & 0.042 & 0.057    & 0.064       & 0.043                                                            & 0.056                                                          & 0.104                      \\ \hline
Avg. \#activated users & 13.99&16.84    & 17.68       & 15.54                                                           & 17.00                                                       & \multicolumn{1}{r|}{23.81} \\ \hline

 \%impacted user       &32.17 &52.54 & 57.44 & 43.44 & 56.54 & 71.66                     \\ \hline
\end{tabular} }
\vspace{-4mm}
\end{table}

 \subsection{Validation of UBM}
 \label{UBM_check}
\paratitle{Experimental results.}
We compare the effectiveness of our UBM to five widely used topology-based measurements
in the literature, i.e., in-degree, PageRank~\cite{page1999pagerank}, TwitterRank~\cite{weng2010twitterrank}, 
betweenness centrality~\cite{freeman1978centrality} and  community centrality~\cite{freeman1978centrality}. 
We randomly split the set of cascades into two subsets. The first subset accounts for 
80\% of the cascades and is used to 
calculate the bridging performance of all users. 
Then we select the top 20\% users with the highest bridging performance in every adopted measurement
and use the other subset to compare their actual influences in information diffusion.
We adopt three measurements to quantitatively assess the effectiveness of UBM and the benchmarks. 
We use the \emph{average number of activated users per minute} to evaluate 
the {efficiency} of the information diffusion. The more users activated in a minute, the faster 
information can be spread when it is shared by the influential users.  
The \emph{average number of activated users} counts the users
who received the information after the retweeting behaviour of an identified influential user.
It is meant to evaluate the expected wideness of the spread once an influential user
retweets a message. 
The \emph{percentage of impacted users} gives the proportion of users that have ever received a 
message due to the sharing behaviours of identified influential users. 
This measurement is to compare the overall accumulated influence of all the selected influential users.  
We show the results of UBM and other benchmark measurements in Table~\ref{tab:comparison}. 
We can observe that it takes less time on average for the influential users
identified according to UBM to activate an additional user, with 0.104 users activated a minute due
to their retweets.  With 23.81 users activated, UBM allows for finding the users
whose retweeting action can reach more than 35\% users than those identified by the benchmarks.
In the end, the top 20\% influential users identified by UBM  spread their shared information 
to 71\% users in our dataset, which overwhelms that of the best benchmark by about 15\%.
From the above analysis in terms of the three measurements, we can see that our UBM can 
successfully identify influential users whose sharing on social media manages to promote 
both the wideness and the speed of the diffusion of COVID-19 related information.

\begin{figure}[t]
\centering
\includegraphics[scale=0.7]{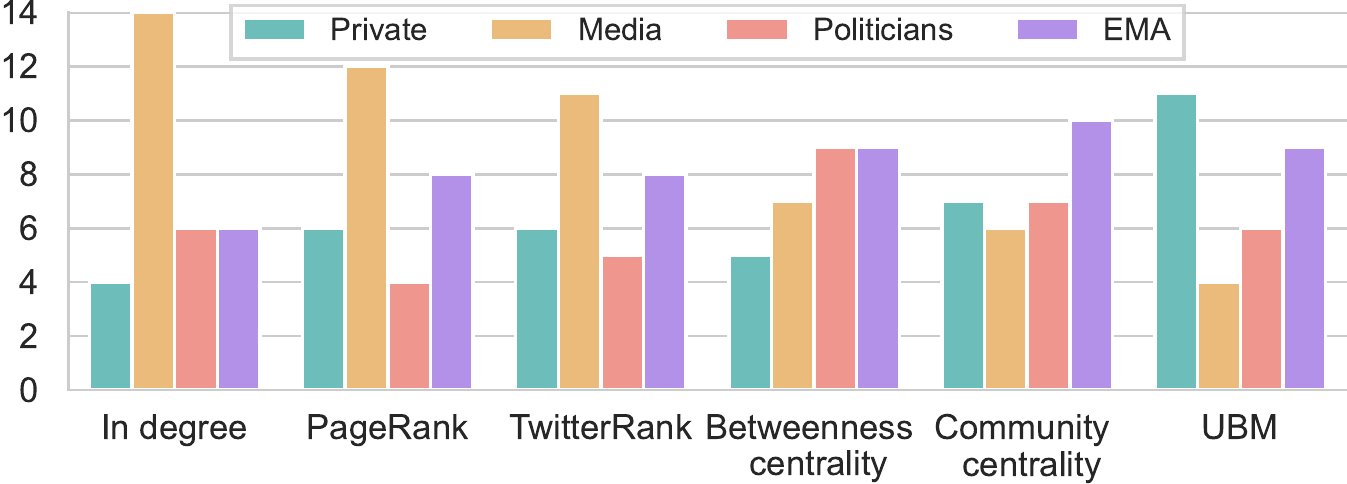}
 \caption{Profile distribution of the top 30 accounts with highest bridging performance}
\label{fig:class}
\vspace{-4mm}
\end{figure}

\paratitle{Manual analysis.} 
In order to understand the profiles of the calculated influential users by the 
measurements, we select the top 30 users with {the} highest bridging performance of 
each measurement. We identify four types of user profiles: \emph{private}, \emph{media}, 
\emph{politicians} and \emph{emergency  management agencies} (EMA).   
Figure~\ref{fig:class} shows the distributions of their profiles.  
We can observe that the distributions vary due to the
different semantics of social connections captured by the measurements. 
For instance, due to the large numbers of followers, Twitter accounts managed by 
traditional media are favoured by in-degree.
This obviously underestimates the importance of accounts such as those of EMAs 
in publishing pandemic updates. 
With reachability and importance in connecting users and communities considered, 
more accounts of politicians and EMAs stand out.
The proportion of private accounts also starts to increase.
When UBM is applied, the percentage of private accounts
becomes dominant. 
A closer check discovers that 10 out of the 11 private accounts belong to health professionals and celebrities.
This is consistent with the literature~\cite{hernandez2021covid} which 
highlights the importance of health professionals and individuals in 
broadcasting useful messages about preventive measures 
and healthcare suggestions in the pandemic.

\section{Impact of COVID-19 on the SWB of Influential Users} 
\label{sec:negativity}
%
In this section, we address the second challenge by 
evaluating individual users' SWB with their textual posts on social media. 
We first study the SWB changes of the users who play a bridging role in transmitting 
COVID-19 related information after the outbreak of the pandemic.
We then analyse whether a user's bridging performance 
relates to his/her SWB changes. 

\subsection{Measuring SWB}

We extend the definition proposed in~\cite{ZJZ20} to measure the level of 
subjective well-being of users based on the sentiment expressed in 
their past tweets. Specifically, we extend it from bi-polarity labels, i.e., negative
and positive affection, to tri-polarity with neutral sentiment by 
multiplying a scaling factor to simulate the trustworthiness of the bi-polarity SWB.
\begin{dfn}[Social media Subjective well-being value (SWB)]
    We use $N_p(u)$, $N_{\it neg}(u)$ and $N_{\it neu}(u)$ to denote the number of positive, 
    negative and neutral posts of a user $u$, respectively. 
    The subjective well-being value of $u$, denoted by $swb(u)$, is calculated as:
    $$
    \frac{N_p(u)-N_{\it neg}(u)}{N_p(u)+N_{\it neg}(u)}\cdot
   \left(\frac{N_p(u)+N_{\it neg}(u)}{N_p(u)+N_{\it neg}(u)+N_{\it neu}(u)}\right)^{\frac{1}{2}}.
    $$
\end{dfn}
\noindent
If all messages are neutral, then ${\it swb}(u)$ is $0$. 

\paratitle{Discussion.}
Note that i) consistent with~\cite{ZJZ20}, we focus on affective SWB 
(i.e., positive and negative) in this paper, while ignoring its cognitive 
dimension; ii) users' SWB is evaluated based on their original messages: 
originally posted tweets and quotations;
iii) for tweets with quotations to other messages, only the texts are 
considered without the quoted messages. 
As retweets may not explicitly include users' subjective opinions, we exclude them 
from the SWB calculation.

\subsection{Analysing SWB changes of influential users}

\begin{figure}[t]
\centering
\includegraphics[scale=0.7]{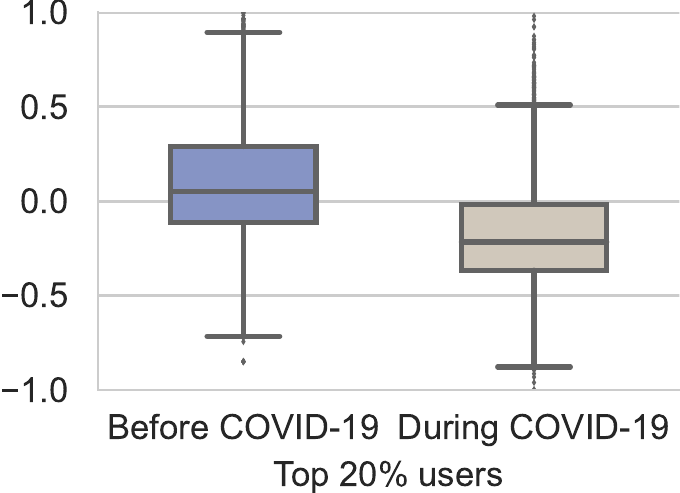}
\hspace{3em}
\includegraphics[scale=0.7]{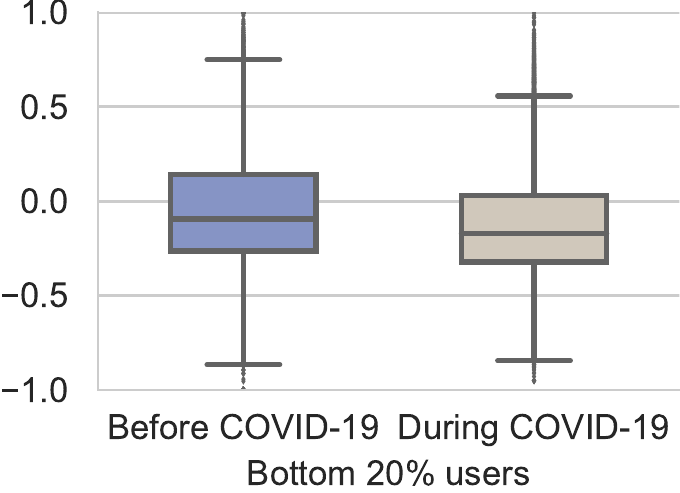}
\caption{SWB changes after the outbreak of the pandemic.} 
\label{fig:box_plot_10}
\vspace{-4mm}
\end{figure}
With the proposed SWB measurement, we study how users' subjective 
well-being changes due to the outbreak of the COVID-19 pandemic. 
We calculate the UBM values of the users in our 
collected dataset and order them descendingly. 
Then we select the top 20\% users as well as the bottom 20\% users 
and compare the two groups' responses to the pandemic.
For each group, we calculate users' SWBs according to their posts 
before the pandemic and after the pandemic to capture the changes.
Note that we only consider the users with more than 5 posts in each 
time period.

In Figure~\ref{fig:box_plot_10}, we show the SWB distributions of the two 
user groups. 
On average, the users with high UBM have positive SWB of 0.11 
before the pandemic while the users with low UBM are negative. 
\emph{All users' SWB decreases after the pandemic but 
the SWB of the top 20\% users drops more significantly.} 
Specifically, their SWB falls by $0.33$, which is two times as much as that of 
the bottom 20\% users. This indicates that the top 20\% users become even 
more negative than the bottom 20\% users.
To sum up, the pandemic causes more negative mental impacts on the social 
media users who play a more important bridging role in 
transmitting COVID-19 related information. 

\subsection{Relation between SWB and bridging performance}

\begin{table}[t]
\centering
 \caption{Hierarchical multiple regression model examining variance in SWB explained  by independent variables, $*p<0.05$; $**p<0.001$}
    \label{tab:correlation}
{\small
\begin{tabular}{lrrrrrrr}
\hline

\textbf{Variable} &
  $B~~~$ & ${\it SEB}$ & $b~~~$ & $t~~~$ & $R~$ & $R^2$ &
  $\Delta R^2$ \\\hline\hline
\textbf{Stage 1}       &        &        &        &        & -0.207 & 0.043 & 0.043 \\
In-degree              & 0.234 & 0.103  & 0.160  & 2.272*  &        &       &       \\
Out-degree             & 0.861  & 0.680  & 0.054  & 1.267  &        &       &       \\
Pagerank               & 3.081  & 0.148  & 0.180  & 2.082*  &        &       &       \\
Betweenness centrality & -3.287 &0.728 & -1.453 & -4.515** &        &       &       \\
\hline
\textbf{Stage 2}       &        &        &        &        & -0.312 & 0.097 & 0.054 \\
In-degree              & 0.228  & 0.102  & 0.158  & 2.239*  &        &       &       \\
Out-degree             & 0.075  & 0.080  & 0.050  & 0.945  &        &       &       \\
Pagerank               &0.307  & 0.150  & 0.180  & 2.049*  &        &       &       \\
Betweenness centrality & -3.268 & 0.723 & -1.390 & -4.520** &        &       &       \\
Activity               & 0.861  & 0.123  & 0.037  & 0.716  &        &       &       \\
\hline
\textbf{Stage 3}       &        &        &        &        & -0.579 & {0.335} & \textbf{0.238} \\
In-degree              & 0.158  & 0.123  & 0.107  & 1.125*  &        &       &       \\
Out-degree             & 0.516  & 0.45  & 0.050  & 1.147  &        &       &       \\
Pagerank             & 0.191  & 0.143  & 0.168  & 1.338*  &        &       &       \\
Betweenness centrality & -1.105 & 0.541 & -0.509 & -2.066** &        &       &       \\
Activity               & 0.067  & 0.133  & 0.053  & 0.508  &        &       &       \\
UBM                    & -2.254 & -0.196 & -1.797  & \textbf{-11.469**}  &        &       &      \\\hline
\end{tabular}
}
\vspace{-5mm}
\end{table}

We conduct the first attempt to study if a user's bridging performance 
has a relationship with the SWB changes of the users actively participating 
the diffusion of COVID-19 related information. 
In addition to UBM and the five benchmark measurements used in Section~\ref{UBM_check}, 
we consider two additional variables: \emph{out-degree} and \emph{activity}. 
Out-degree is used to check whether the number of accounts a user follows 
correlates with SWB changes. The activity variable evaluates how active 
a user is engaged in the online discourse and is quantified by the number of 
messages he/she posted.  
In order to isolate the impacts of these variables, we adopt the method of 
\emph{hierarchical multiple regression}~\cite{tabachnick2007using}. 
The intuitive idea is to check whether the variables of interest can 
explain the SWB variance  after accounting for some variables. 

To check the validity of applying hierarchical multiple regression,
we conduct first-line tests to ensure a sufficiently large sample size and 
independence between variables. We identify the variables corresponding to 
community centrality  and TwitterRank fail to satisfy the multi-collinearity requirement. 
We thus ignore them in our analysis. 
The ratio of the number of variables to the sample size is 1:2108, which is well below the 
requirement of 1:15~\cite{tabachnick2007using}. This indicates the sample size is adequate. 
We iteratively input the variables into the model with three stages.  
The results are shown in Table~\ref{tab:correlation}.
In the first stage, we input the variables related to network structures, i.e.,  
in-degree, out-degree, Pagerank and Betweenness centrality. 
The combination of the variables can explain $4.30\%$ of the SWB variance 
($F=4.379, p<0.05$). 
Note that an F-value of greater than 4 indicates the linear equation can 
explain the relation between SWB and the variables. 
This demonstrates that there exists a positive relationship between the 
topology-based variables and SWB, but this relationship is rather weak. 
A closer check on  the $t$-values show that 
out-degree is irrelevant to SWB and the rest three variables are weakly related.  
In the second stage, we add the variable of activity to the model. 
After controlling all the variables of the first stage, we 
observe that user activity does not significantly contribute to the 
model with $t$-value of $0.716$. 
This suggests that user activity is not a predictor of SWB.
In the third stage, we introduce UBM to the model. 
The addition of UBM, with the variables in the previous two stages controlled, 
reduces the R value from -0.219 to -0.579. 
UBM contributes significantly to the overall model with $ F=147.82$ ($p < 0.001$) 
and increases the predicted SWB variance by 23.8\%.
Together with the $t$-value of -11.469 ($p<0.001$), we can see 
there exists a strong negative relation between UBM and SWB, and 
UBM is a strong predictor for SWB.

\paratitle{Discussion.}
To conclude, the results illustrate that UBM is strongly related to SWB, while 
in-degree, Pagerank and betweenness centrality are weakly related.
This difference further shows that UBM can more accurately capture users’ behaviour 
changes after the outbreak  of the pandemic while topology features remain similar 
to those before the pandemic.
This may be explained by the recent studies~\cite{HLCX20} that  
once considered as a change in life after the pandemic outbreak, 
this extra bridging responsibility in 
diffusing COVID-19 related messages is likely to associate with lower life satisfaction.

\section{Conclusion and Limitation}
\label{sec:conclusion}
In this paper, we concentrated on the social media users whose sharing behaviours 
significantly promote the popularity of COVID-19 related messages. 
By proposing a new measurement for bridging performance,  we identified these 
influential users. With our collected Twitter data of an international region,  
we successfully show the influential users suffer from more decrease in 
their subjective well-being compared to those with smaller bridging performance. 
We then conducted the first research to reveal the strong {relationship} between 
a user's bridging performance in COVID-19 information diffusion 
and his/her SWB.  
Our research provides a cautious reference to public health bodies 
that some users can be mobilised to help spread health information, 
but special attention should be paid to their psychological health. 

\smallskip\noindent \textbf{Limitations and our future work.}
This paper has a few limitations that deserve further discussion. 
First, we only focused on the affective dimension of subjective 
well-being while noticing its multi-dimensional nature. This allows us to follow
previous SWB studies to convert the calculation of SWB to sentiment analysis, 
but does not comprehensively evaluate users' cognitive well-being, such as life satisfaction. 
In our following research,  we will attempt to leverage more advanced AI models to 
investigate cognitive aspects such as \emph{happy} and \emph{angry}.  
Second, extracting SWB from users' online disclosure inevitably incurs bias 
compared to social surveys
although it supports analysis of an unprecedented large number of users. 
Last, we notice that the region we targeted at may introduce additional bias 
in our results.  
As a continuous work, we will extend our study to a region of multiple European 
countries and cross-validate our findings with other published results in social science.

\paratitle{Ethical considerations.}
This work is based completely on public data and does not contain private 
information of individuals. Our dataset is built in accordance with the FAIR 
data principles~\cite{wilkinson2016fair} and Twitter Developer Agreement and Policy and related policies. 
Meanwhile, there have been a significant amount of studies on measuring 
users' subjective well-being through social media data. 
It has become a consensus that following the terms of service of 
social media networks is adequate to respect users' privacy
in research~\cite{fernando2020towards}. 
To conclude, we have no ethical violation in the collection and 
interpretation of data in our study.
\bibliographystyle{splncs04}
\bibliography{ref}

\end{document}